\documentclass[9pt,preprint2,natbib209]{aastex}

\usepackage{graphicx}
\usepackage{dcolumn}
\usepackage{bm}
\shorttitle{ The Shoenberg Effect in Magnetars} \shortauthors{Wang
et al.}

\begin{document}

\title{ The Shoenberg Effect in Relativistic Degenerate Electron Gas and Observational Evidences in Magnetars }

\author{Zhaojun Wang\altaffilmark{1}, Guoliang L\"{u}\altaffilmark{1}, Chunhua Zhu\altaffilmark{1}, Wensheng Huo\altaffilmark{1}}
\email{xjdxwzj@sohu.com, guolianglv@gmail.com}
\altaffiltext{1}{School of Physical Science and Technology, Xinjiang
University, Urumqi, 830046, China}

\begin{abstract}
The electron gas inside a neutron star is highly degenerate and relativistic.
Due to the electron-electron magnetic interaction, the differential
susceptibility can equal or exceed 1, which causes  the magnetic
system of the neutron star to become metastable or unstable. The
Fermi liquid of nucleons under the crust can be in a metastable state,
while the crust is unstable to
the formation of layers of  alternating magnetization. The
change of the magnetic stress acting on adjacent domains can result
in a series of shifts or fractures in the crust. The releasing of
magnetic free energy and elastic energy in the crust can
cause the bursts observed in magnetars. Simultaneously, a series of
shifts or fractures in the deep crust which is closed to the Fermi
liquid of nucleons can trigger the phase transition of the Fermi
liquid of nucleons from a metastable state to a stable state. The
released magnetic free energy in the Fermi liquid of nucleons
corresponds to the giant flares observed in some magnetars.

\end{abstract}

\maketitle

\section{Introduction}

In 1930, de Haas and van Alphen first observed that the
magnetization $\vec{M}$ in normal metals oscillates  in low temperature,
under an applied intense magnetic field. The oscillatory
functions are sinusoidal series with the fundamental frequency that
can be described by the extremal areas of the cross-section of the
Fermi surface normal to the applied magnetic field
\citep{lifshitz1956}. Using the impulsive field method, Shoenberg
found an unexpectedly high amplitude for the second harmonic, and
proposed the magnetic interaction among the conduction electrons
\citep{Shoenberg1962}, the so-called Shoenberg effect.
He suggested that the magnetizing field is not the applied field
$\vec{H}$ but the magnetic induction $\vec{B}$. When the differential
magnetic susceptibility $\chi_{\rm m}=\partial(4\pi M)/\partial B$
exceeds 1 ( We use the Gaussian units in the paper), the
spatially uniform state of an electron gas is thermodynamically
unstable. Then, the electron gas rapidly evolves into a stable
state, and has    a spatially inhomogeneous magnetic field with
 various Condon domains \citep{condon1966}.

The magnetization and the de Haas-van Alphen oscillating effect for
a relativistic degenerate electron gas have been studied by many
authors \citep{Visvanathan62,canuto68,chudnovsky81}. Previous
studies about the magnetic susceptibility of neutron stars mainly focused
on whether or not the observed field is resulted from a spontaneous
magnetization \citep{lee69,oconnell71}. This almost cannot occur
because the neutron star is insufficiently cool. Blandford and Wilkes
\citep{Blandford82,Wilkes89} proposed the domain
structure and applied it to  neutron star crusts. However, the inhomogeneous field
resulted from domain formations in the crust of a normal neutron
star is negligible and can not produce observable effects. On the other hand, for
magnetars with magnetic fields stronger than normal neutron stars,
the inhomogeneous field is large enough  to produce
super-Eddington X-ray outbursts.

Magnetars, including Anomalous X-ray Pulsars (AXPs) and Soft
Gamma Repeaters (SGRs), are characterized by their inferred dipolar
magnetic field strength ranging from $5.9\times10^{13}$ \mbox{G} to
$1.8\times10^{15}$ \mbox{G}, and their long spin periods ranging from
5.2 \mbox{s} to 11.8 \mbox{s}.  Surprisingly, some magnetars have
undergone giant flares (GFs) in which  the energy up to $10^{46}$
ergs is released in a fraction of a second via $\gamma$-ray
emissions. Up to now, three GFs have been observed, respectively, from
SGR 0526-66\citep{Cline82}, SGR 1900+14\citep{Mazets99} and SGR
1806-20 \citep{Palmer05}.

Their X-ray and $\gamma$-ray
luminosities during  persistent and burst phases are too large to be
powered by their kinetic energy.
Where is the released magnetic energy  stored prior to the GFs? Is it
in the magnetospheres or in the neutron star? \cite{Duncan92} and
\cite{TD01} proposed a  model for magnetars, and
considered that the released magnetic energy is stored in the
neutron star. The controversy of this model is about its
triggering mechanism for high energy radiation. \cite{Duncan92}
suggested that a helical distortion of the magnetic field in the
core induces a large-scale fracture in the crust and a twisting
deformation of the magnetic field in the crust and
magnetospheres. A GF may involve a large disturbance which
probably is driven by a rearrangement of the magnetic field in the deep
crust and core \citep{TD95}, while the persistent emission can be
explained by the very slow transport of the field from the core to
the crust by the Hall drift \citep{Goldreich92}. \cite{Kondratyev02}
first postulated the existence of magnetic domains in the magnetar
crust because of the inhomogeneous crust structure. He argued that
the burst activity of SGRs could originate from magnetic avalanches.
However, because of the domain forming mechanism resulted from the
interaction among nucleus spins, the required magnetic field must be in
the range $10^{16}\sim 10^{17}$ G. It is still an open question whether or not
the magnetic field in the crust of a neutron star can be so strong.
Having considered the GF's emission energy
($10^{42}-10^{46}$ \mbox{ergs})  and its mean waiting time
\citep{Mazets79,Hurley99,Terasawa05}, \cite{Stella05} estimated that
the internal field strength of a magnetar at its birth time can reach up
to $B\geq10^{15.7}$ \mbox{G}. Later, after numerically calculating
the magnetic properties of magnetar-matter,  such as the magnetization
and  susceptibility of electrons, protons and neutrons, \cite{Suh10}
proposed a magnetic domain model to correlate smoothly between the
statistics of star-quakes and the magnetic avalanches in the magnetar
crust.

The phase transition to the domain phase in magnetars is
different from the ferromagnetic phase transition. The magnetic
ordering in iron comes from the exchange interaction of bound
electron spins at a sufficient low temperature, and it is not
prerequisite to have an external magnetic field. While the magnetic
ordering in magnetars arises from the interaction among the orbital
magnetic moments of free electrons under high-quantizing field.
Therefore, given  a nearly uniform distribution of electrons, the
phase transitions in magnetars are similar to those in beryllium
where magnetized matter is the conductive
electrons \citep{Shoenberg1962}. The relativistic Fermi sea of
electrons is only slightly perturbed by the Coulomb forces of nuclei
in the crust of the magnetar and does not efficiently screen nuclear
charges. The dominant contribution to the differential susceptibility in
a magnetar comes from electrons, while the contributions from
nucleons are negligible (the magnetic moments of nucleons are three
orders lower than the electron orbital moments).

In this paper,  we analytically calculate the differential
susceptibility of a relativistic degenerate electron gas, and find that it
is an oscillating function of the magnetic field,  which is often called the
de Haas and van Alphen oscillation. In \S 2, we discuss the magnetic
phase transition, while in \S 3 we calculate the differential susceptibility.
In \S 4 we use our model to explain the observed evidences
in magnetars, and finally summarize our main results in \S 5.

\section{The magnetic phase transition}

 The magnetization of the degenerate electron gas in a
highly-quantizing magnetic field and low temperature ($KT\ll \hbar
\omega_{\rm c}$, where $\omega_{\rm c}$ is the electron cyclotron
frequency in magnetic field) should exhibit a nonlinear de Haas-van
Alphen effect. Without the magnetic interaction the oscillating
magnetization is assumed to be periodical in magnetic field $H$
rather than in $1/H$ \citep{Pippard80,Shoenberg84}:
\begin{equation}
\tilde{M}=M_{\rm 0}\sin(H/H_{\rm 0}),
\end{equation}
where $H_{\rm 0}$ hardly varies over one cycle of oscillation.
However, if we consider the feedback contribution from
$4\pi\tilde{M}$, the magnetization depends on not only the
quantizing magnetic field but also the cooperating ordering of
the magnetic moments. Thus, the magnetization  should actually
be a function of the magnetic induction, $\vec{M}=\vec{M}(\vec{B})$.
Then, the magnetic induction is given by $\vec{B}=\vec{H}+4\pi
\vec{M}(\vec{B})$. Replacing $\vec{H}$ by $\vec{B}$ is called  the
Shoenberg or $\vec{H}$-$\vec{B}$ effect. The oscillating sinusoidal
function of the magnetization in the magnetic field of Eq.(1) is given
by
\begin{equation}
\tilde{M}=M_{\rm 0}\sin(B/H_{\rm 0}) =M_{\rm 0}\sin[(H+4\pi
M)/H_{\rm 0}],
\end{equation}
where $M_{\rm 0}$ and $H_{\rm 0}$ are constants.

When $4\pi M_{0}/H_{0}>1$, as Fig.1 shows, $\tilde{M}$
becomes a three-values function of $H$, and the differential
magnetic susceptibility $\chi_{\rm m}=\partial(4\pi
\tilde{M})/\partial B$ can exceed one. Considering  a thin rod oriented along the direction of the
magnetization,
\cite{pippard1963} found that the multi-valued function in Fig.1 is not
physical. In the region of the curve between L and $\mbox{L}'$ where
the slope is lower than one, the magnetized states are unstable
and can never really exist. For any weak perturbation $\delta
\vec{H}$ near $\vec{H}_{0}$, the perturbation of magnetic induction
$\delta \vec{B}$ is given by
\begin{equation}
\delta\vec{B}=\frac{\delta\vec{H}}{1-\chi_{\rm m}}\mbox{.}
\end{equation}
Obviously, there is a singular point when $\chi_{\rm m}=1$ in
Eq.(3), where the magnetized state is unstable and the first-order
phase transition should occur. Then, the magnetic system should be in
a stable state in which the magnetization is inhomogeneous and magnetic domains
form. The magnetizations in the adjacent domains have opposite directions.
The stable magnetized states
are represented by the dashed line between N and $\mbox{N}'$ in
Fig.1. But if there is a surface energy at the boundary of two
different magnetizations it may become metastable, which is similar
to superheating or supercooling in gas-liquid transition
\citep{Reichl98}. The solid line between L and N or $\mbox{L}'$ and
$\mbox{N}'$ in Fig.1 represents the metastable state. It is not
difficult to achieve the condition of the above magnetic phase transition in the terrestrial
labs.  The experiments of Condon indicated that the magnetized phase
transition can take place in some metals \citep{condon1966}.
However, what is the situation in compact objects such as neutron
stars? In the following we calculate the differential susceptibility
of a relativistic degenerate electron gas and show the observable
effects of the magnetic phase transition in neutron stars.
\begin{figure}
  \includegraphics[width=8cm]{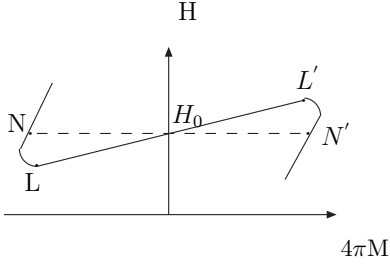}\\
  \caption{Magnetic field $H$ vs. magnetization $4\pi \tilde{M}$}\label{b}
\end{figure}\\

\section{The differential susceptibility of a relativistic degenerate electron gas}

The assembly of electrons in a neutron star under a strong
magnetic field is degenerated and relativistic. The energy
eigenvalues are \citep{Jonhson49}
\begin{equation}E=[c^{2}p_{\rm z}^{2}+\mu^{2}+\mu \epsilon_{\rm c}(2n+s+1)]^{1/2}\mbox{ ,}
\end{equation}
where $\mu=m_{\rm e}c^{2}$ and $\epsilon_{\rm c}=\hbar eB/cm_{\rm
e}=\hbar \omega_{\rm c}$ are the rest energy and cyclotron energy of
electron, respectively. Here, $p_{\rm z}$ is the momentum component
along the field direction which is taken as the z-direction, $n=0,1,... $, and
$s=\pm 1$ are the Landau and spin quantum number, respectively. The
density of states per unit volume for the energy level is given by
$(p_{\rm z}/h)(eB/hc)=g(p_{\rm z}/h)$,  where $g=eB/hc$ is the density
of states per unit area for a Landau energy level. As a consequence
of quantization, the spherical Fermi surface of the free electron is
replaced by a set of circles located on a spherical surface with a
common axis along the $B$ direction, which is shown in Fig.2.
\begin{figure}
  \includegraphics[width=5cm]{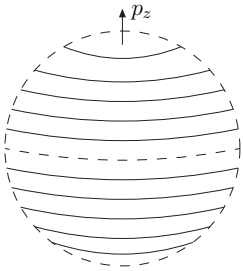}\\
  \caption{The free-electron Fermi surface in the presence of a magnetic field
along the $p_{z}$-axis.}\label{b}
\end{figure}

 The susceptibility of the electron assembly can be
obtained by finding its grand potential which depends on the density
of states. It is a function of energy and magnetic induction. The
density of states per unit volume is given by
\begin{equation}
Z(\epsilon,B)=\frac{2eB}{c^{2}h^{2}}\sum_{n,s}
[\epsilon^{2}+2\epsilon\mu-\mu \epsilon_{c}(2n+s+1)]^{1/2}\mbox{ ,}
\end{equation}
where $\epsilon=E-\mu$ is the kinetic energy of an electron. The
grand potential per unit volume of the assembly is given by
\begin{equation}
\begin{array}{ll}
J&=-\beta \int
\ln[1+e^{\beta(\psi-\epsilon)}]\mathrm{d}Z(\epsilon,B)\nonumber\\
&=-\int_{0}^\infty Z(\epsilon,B) f(\epsilon)\mathrm{d}\epsilon
\mbox{ ,}
\end{array}
\end{equation}
where $\beta=(kT)^{-1}$, $\psi$ is the chemical potential, and
$f(\epsilon)$ is the Fermi-Dirac distribution function. After
summing over the spin quantum number, Eq.(5) reduces to
\begin{equation}Z=\frac{2}{\sqrt{\pi}}g^{3/2} [b^{1/2}+2\sum_{n=1}^{[b]} (b-n)^{1/2}]
\mbox{,}
\end{equation}
where
\begin{equation}b=\frac{\epsilon^{2}+2\epsilon \mu}{2\epsilon_{\rm c}\mu}\mbox{
,}
\end{equation}
and $[b]$ is the integer of $b$ ($[b]\leq b$). For a neutron star,
the electron system  is almost completely degenerate ($\psi\gg
\beta^{-1}$) and the Fermi-Dirac distribution function is almost a
step function except the region near $\epsilon=\psi$. The magnetic
moment depends on the first-order derivative of the grand potential
with $B$, and it mainly comes from the first-order derivative of Eq.(6)
at $\epsilon=\psi$ Hence, we define $b_{m}=b|_{\epsilon=\psi}$. If the
chemical potential of the neutron star is $\sim$ 10 MeV and the
magnetic field is the quantum field $B_{\rm Q}=4.414\times 10^{13}$G
(the cyclotron energy of electron equals to its rest energy), then
$[b_{m}]$ is about 200. Based on Eq.(7), we can see that the
first-order derivative of the grand potential  has a singularity
when $b_{m}=[b_{m}]$. Therefore, the susceptibility of the
relativistic degenerate electron gas can be equal to or even larger
than 1.

To reveal the oscillating effect of the grand potential, we
calculate the sum of Eq.(7) with the Poisson summation formula
\citep{dingle1952}:
\begin{equation}\frac{1}{2}F(0)+\sum_{n=1}^\infty F(n)=\sum_{r=-\infty}^\infty
\int_0^{[b_{\rm m}]} F(x)e^{i2{\pi}rx}\mathrm{d}x\mbox{.}
\end{equation} Because $[b_{\rm m}]$ is much larger than 1, the density of
states in Eq.(5) is approximately by
\begin{equation}
\begin{array}{ll}Z=&\frac{4}{\sqrt{\pi}}g^{3/2} [\frac{2}{3}b_{\rm m}^{3/2}+\sum_{\nu=0,2,4}^\infty\frac{(-1)^{\nu/2}(2\nu-1)!B_{{(\nu+2)}/2}}
{\sqrt{b}(4b)^{\nu}(\nu-1)!(\nu+2)!}+\nonumber\\
 &\frac{1}{2\sqrt{2}\pi}\sum_{r=1}^\infty \frac{\cos(2\pi rb-3\pi/4)}{r^{3/2}}] \mbox{ ,}\\ \end{array}
 \end{equation}
where $B_{\rm m}$ denotes the Bernoulli numbers. The first two terms
on the right-hand side of the above equation are the non-oscillating parts while the third
term is the oscillating
part.\\

The non-oscillating grand potential per unit volume can be
obtained from Eq.(10). As the first-order approximation, we can keep
only the first term in the summation and obtain
\begin{equation}
\begin{array}{ll}\bar{J}=&-\frac{4}{3\sqrt{\pi}(2 \mu \epsilon_{\rm c})^{3/2}}g^{3/2}
\int_{0}^\infty(\epsilon^{2}+2\epsilon\mu)^{3/2}f(\epsilon)\mathrm{d}\epsilon+\nonumber\\
 &\frac{\sqrt{2 \mu \epsilon_{\rm c}}}{6\sqrt{\pi}}g^{3/2}\int_{0}^\infty\frac{f(\epsilon)\mathrm{d}\epsilon}
 {(\epsilon^{2}+2\epsilon\mu)^{1/2}} \mbox{ .}\end{array}
\end{equation}
In general, $\psi\gg\mu$ for a neutron star. Using the standard
methods of evaluating integrals for the Fermi-Dirac distribution, we
can evaluate the integrals in Eq.(11). The non-oscillating
susceptibility is given by
\begin{equation}
\begin{array}{ll}\bar{\chi}_{\rm m}&=4\pi\frac{\partial^{2} \bar{J}}{\partial B^{2}}\nonumber\\
 &=\frac{4\sqrt{2\pi}}{3}\frac{\sqrt{\mu \epsilon_{\rm c}}}{B^{2}}g^{3/2}
 [\ln \frac{2\psi}{\mu}-\frac{\pi ^{2}}{6}(\frac{kT}{\psi})^{2}] \mbox{.}\end{array}
 \end{equation}
In our work, the chemical potential and the magnetic field in the
deep crust of a normal neutron star are, respectively,  about 10 MeV and $10^{12}$ G,
while the non-oscillating susceptibility can be
$\sim 5.6\times 10^{-3}$.\\

Using the approximate formula for the Fermi-Dirac distribution
\begin{equation}\int_{0}^\infty \eta
(\epsilon)f(\epsilon)\mathrm{d}\epsilon= \int_{0}^{\psi-\mu}\eta
(\epsilon)\mathrm{d}\epsilon+\frac{\pi^{2}}{6}(kT)^{2}\eta'(\psi-\mu)\mathrm{d}\epsilon
\mbox{ ,}
\end{equation}
and the Fresnel Integral
\begin{equation}
\begin{array}{lll}\int_{0}^x\cos(\frac{\pi}{2}t^{2})\mathrm{d}t&\approx
 x \ \ \ \ (x\ll1)\nonumber\\
 & \approx \frac{1}{2}+\frac{1}{\pi x}\sin(\frac{\pi}{2}x^{2}) &
(x\gg1)\mbox{ ,}\end{array}
\end{equation}
we  obtain the oscillating term of the grand potential per unit
volume,
\begin{equation}
\begin{array}{ll}\tilde{J}=&\frac{\sqrt 2}{\pi^{3/2}}g^{3/2}\sum_{r=1}^\infty
\frac{1}{r^{3/2}}[(\frac{1}{2}\sqrt
\frac{\pi}{2a_{\rm r}}-\mu)\cos(a_{\rm r}\mu^{2}+\frac{3}{4}\pi)+\nonumber\\
&\frac{1}{2}\sqrt
\frac{\pi}{2a_{\rm r}}\sin(a_{\rm r}\mu^{2}+\frac{3}{4}\pi)]+\frac{\sqrt 2}{\pi^{3/2}}g^{3/2}\sum_{r=1}^\infty\frac{1}{r^{3/2}}\cdot \nonumber\\
&\frac{1}{a_{\rm
r}\psi}[\frac{\pi}{2}-\frac{\pi^{2}}{3}(kT)^{2}(a_{\rm
r}\psi)^{2}]\sin(a_{\rm r}\psi^{2}-a_{\rm
r}\mu^{2}-\frac{3}{4}\pi)]\mbox{ ,}\end{array}
\end{equation}
where $a_{\rm r}=\pi r/(\epsilon_{\rm c}\mu)$. When calculating the
differential susceptibility which is the second derivative of
Eq.(15), we only kept the most rapidly varying terms, that is, only
differentiating cosines and sinusoidal. Because $\psi\gg \mu$ for
a neutron star, the last term in Eq.(15) is dominant. The oscillating
susceptibility is approximately given by
\begin{equation}
\tilde{\chi}_{\rm m}=A_{0}\sum_{r=1}^\infty (r^{-1/2}-A_{\rm
1}(\frac{kT}{\epsilon_{\rm c}})^{2}r^{3/2})\cos(a_{\rm
r}\psi^{2}-\frac{3}{4}\pi)\mbox{,}
\end{equation}
where
\begin{equation}
A_{0}=\alpha(\frac{\psi}{\mu})^{3}(\frac{B}{B_{\rm Q}})^{-3/2}
\mbox{,}
\end{equation} and
\begin{equation}
A_{1}=\frac{2 \pi^{3}}{3}(\frac{\psi}{\mu})^{2}\mbox{.}
\end{equation}
Here, $\alpha$ in Eq.(17) is the fine structure constant. The result
of Eq.(16) is similar to the result obtained in \cite{Visvanathan62}. It
indicates that the susceptibility oscillates with $1/B$. We replace
$B$ by $B^\prime=B-B_0$. When $B\rightarrow B_0$ we have
$1/B\approx1/B_0(1-B^\prime/B_0)$. Note that the first term is a
constant, thus the oscillation is a period function of $B'$. The
differences of the magnetic induction in an oscillating period are the
periods of $B'$, which are given by
\begin{equation}
\delta B=\frac{2\hbar ec^{2}B_{0}^{2}}{\psi^{2}r}\ \ \ \ \ \ \ \ \ \
\ \ \ \ \ (r=1,2,\cdots)\mbox{.}
\end{equation}
The coefficients of the cosine functions in Eq.(16) include two
terms. The first term is the result of complete degeneracy, while
the second term comes from the thermal fluctuation and is
proportional to the second power of temperature. For a typical
neutron star, the chemical potential, magnetic field and temperature are
of the order of 10 MeV, $10^{12}$ G, and $10^{6}$ K, respectively. The
second term can be ignored because of $KT\ll \hbar \omega_{\rm c}
\ll \psi_{\rm 0}$. The oscillating susceptibility,
$\tilde{\chi}_{\rm m}$, is mainly determined by $A_{\rm 0}$. Fig.3
shows the relation between $\log A_0$ and $B/B_{\rm Q}$. Obviously,
the differential susceptibility can equal or exceed 1 when $B<\sim15
B_{\rm Q}$. In this work, we assume that the magnetars are similar to
the normal neutron stars except their magnetic fields: for a normal neutron
star, $B=10^{12}$ G,  while for a magnetar $B=15B_{\rm Q}$. For a
stronger magnetic field, the condition $\hbar \omega_{\rm c} \ll
\psi_{\rm 0}$ or $[b_{\rm m}]\gg 1$ cannot be satisfied, and the above
approximate methods cannot be used. We will discuss them in the next
paper.

\begin{figure}
  \includegraphics[totalheight=6cm,width=6cm,angle=-90]{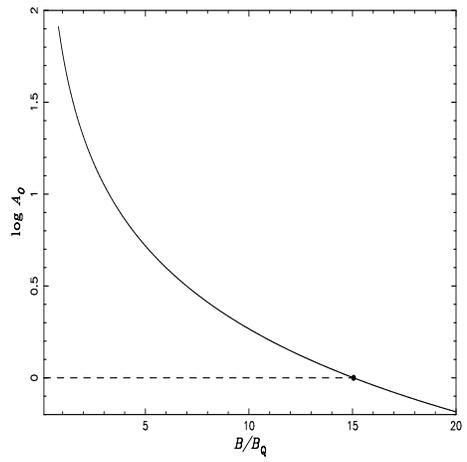}\\
  \caption{$\log A_{0}$ vs. $B/B_{\rm Q}$}\label{b}
\end{figure}

\section{The observed evidences in magnetars}

As mentioned in the last section, the susceptibility of a
relativistic degenerate  electron gas can be equal to or exceed 1.
The electron gas under a strong magnetic field in a neutron star may be in a
unstable
state, and the first-order phase transition should occur. Finally, the electron
gas should be a stable state and has the Condon domain
structure. The magnetizations
in the adjacent domains have opposite directions.

Based on Eq.(19), the relative variation of the magnetic induction
between adjacent magnetic domains is
\begin{equation}
\frac{\delta B}{B_{\rm 0}}\sim 10^{-3}\frac{B_{\rm 0}}{B_{\rm Q}}.
\end{equation}
For a normal neutron star ($B_{\rm 0}\sim 10^{12}$ G), $\delta
B/B_{\rm 0}\sim10^{-4}$, it is very difficult to observe the
oscillatory effects. However, for magnetars ($B_{\rm 0}\sim 15B_{\rm
Q}$), $\delta B/B_{\rm 0}\sim10^{-2}$, which is large enough to
produce the bursts in SGRs or AXPs.

The adjacent magnetic domains have different electron densities. The different magnetizations in the adjacent magnetic
domains mean that the phase difference is $\pi$. According to
Eq.(16), we have $\delta (a_{\rm r}\psi^{2})=\pi$. For simplicity we
approximate the chemical potential $\psi$ with the zero-temperature
free field Fermi energy $\psi_{0}=ch(3n/8\pi)^{1/3}$. The difference
of electron densities between the adjacent magnetic domains is given
by
\begin{equation}
\frac{\delta n}{n}\sim 10^{-2}(\frac{\epsilon_{\rm
e}}{20})^{-2}(\frac{B_0}{15B_{\rm Q}}),
\end{equation}
where $\epsilon_{\rm e}=\psi_{0}/\mu$ and we have taken
$B=B_0$. For a magnetar, $\psi_{\rm 0}\sim 10$ MeV, $B_{\rm 0}\sim
15B_{\rm Q}$, and the rest energy of electrons $\mu\sim 0.5$ MeV.
Obviously, here $\epsilon_{\rm e}$ is scaled  to 20 and $B_{\rm 0}$
to $15B_{\rm Q}$, respectively.

The electrons in a neutron star coexist with other components such
as the nuclei in the crust or the Fermi liquid of protons and neutrons
closed to the deep crust. Due to the Coulomb interactions among
electrons,  nuclei or protons,  the matter of the neutron star may not
be homogeneous,  once the domain structure appears in the crust.
However, it can not occur in the Fermi liquid. The mechanism of
forming electron domains is the interactions among orbital magnetic
moments which are counterbalanced by the degenerate pressure of
electrons $P_{\rm e}$. Because the Coulomb interaction of the
crystal lattices in the crust is the same order of magnitude with
the interaction among the orbital magnetic moments, the electron
domain structure can lead to the formation of the nuclei domain
structure. However, in the Fermi liquid of nucleons, the neutron
degenerate pressure $P_{\rm n}\gg P_{\rm e}=Y_{\rm e}P_{\rm n}$,
where $Y_{\rm e}$ is the fraction of electrons per neutron and it is
only several tenths. The  interaction among the orbital magnetic
moments can not counterbalance  the neutron degenerate
pressure $P_{\rm n}$. Therefore, the magnetic domain structure
cannot form in a Fermi liquid of nucleons. The magnetic system is in a
metastable state, which is similar to the supercooling or
superheating states in the first-order phase transition.

In the crust of a magnetar we can roughly calculate the size of the
domain structure. With the density increasing, the quantum number of
the Landau energy also increases. Using the gravitational potential
energy of a nucleon and the Fermi energy per electron, we can
estimate the height of the magnetic domains as \citep{Suh10}
\begin{equation}
\delta z \sim 10^{3}(\frac{g}{10^{14}})^{-1}(\frac{Y_{\rm
e}}{0.36})(\frac{\epsilon_{\rm e}}{20})^{-1}(\frac{B_{\rm
0}}{15B_{\rm Q}}) \ \ \ \ {\rm cm},
\end{equation}
where $g$ is the surface gravity and $Y_{\rm e}$ is scaled to 0.36
when neutrons drip out from nuclei. The actual size and shape of the
domains is problematical. There seem to be two possibilities
\citep{Blandford82}. The first is that the domains have a horizontal
scale $\delta z$ which is in general required if there is to be a
balance in the electron pressures. The second possibility is that
the domains form a two-dimensional lattice of vertical needles with a
thickness given roughly by the geometrical mean of the cyclonic
radius. This second configuration minimizes the magnetic and
surface energies. If we assume that the volume change of the electron
gas during forming the second domain configuration is only along the
horizontal direction and the changing magnitude (or the thickness of
domain walls) is $\sim$ a cyclonic radius of an electron, based on
Eq.(21), the width of the magnetic domain for the second possibility can
be estimated as
\begin{equation}
\delta l\sim 10^{-9}(\frac{\epsilon_{\rm e}}{20})^{3}(\frac{B_{\rm
0}}{15B_{\rm Q}})^{-2}\ \ \  \ {\rm cm}.
\end{equation}

The Maxwell shearing stress between the adjacent domains can deform
the crust and gives rise to  a strain, $\theta$. It is given by
\begin{equation}
\frac{B_{0}\delta B}{4\pi}\sim \theta \nu \sim \frac{\theta B_{\rm
\nu}^{2}}{4\pi},
\end{equation}
where $\nu$ is the shear modulus of the crust and $B_{\rm
\nu}=\sqrt{4\pi \nu}$. Closed to the bottom of the crust, $B_{\rm
\nu}\simeq 6\times 10^{15}$ G \citep{Baym71}. For magnetars,
$B_{0}\sim 15B_{\rm Q}$, according to Eqs.(20) and (24), we have the
strain $\theta \sim 10^{-3}$. \cite{Ruderman91} found that the
maximum strain of the crust, $\theta_{\rm max}$, is $\sim
10^{-2}-10^{-4}$. Our result clearly is within  the range. When the domain
structures appear in the crust of a magnetar, the shearing stress
acting on adjacent domains can produce a relative shift or a
fracture between them. This sudden shift or fracture can propagate
with the Alfv\'{e}n velocity along the domain layers, which results in a
series of shifts or fractures in the magnetic domains. The time scale of
shifts propagating along a domain layer is estimated as
\begin{equation}
\tau \sim \frac{l}{V_{\rm A}}\sim 0.1(\frac{B_{\rm 0}}{15B_{\rm
Q}})^{-1}(\frac{\rho}{10^{15}})(\frac{l}{R_{\rm
*}})\ \ \  \ {\rm s},
\end{equation}
where $V_{\rm A}$ and $\rho$ are the Alfv\'{e}n velocity
\citep{Duncan04} and the matter density at the deep crust,
respectively, $l$ is the length of the domain layer,  and $R_{\rm
*}$ ($\sim 10 {\rm km}$)  the radius of the magnetar. The time scale
agrees with the observed duration of the SGRs bursts \citep{TD95}.
Available free energy equals the change of magnetic energy from a
domain structure to a homogeneous structure. The density of magnetic free energy is
\begin{equation}
w=\frac{1}{16\pi}[(\vec{B}+\delta\vec{B})^{2}+(\vec{B}-\delta\vec{B})^{2}-2(\vec{B})^{2}]=\frac{(\delta
B)^{2}}{8\pi}.
\end{equation}
Then, the total free energy approximates to
\begin{equation}
E_{\rm burst}\sim \frac{(\delta B)^{2}}{8\pi}4\pi (R_{*})^{2}\delta
z.
\end{equation}
According to Eq.(20) and Eq.(22), the total free energy
is
\begin{equation}
E_{\rm burst}\sim 10^{41}(\frac{\delta B}{10^{13}})^{2}(\frac{\delta
z}{10^{3}}) \ \ \  \ {\rm ergs},
\end{equation}
where we scaled $\delta z$ to $10^{3}$ cm and $\delta B$ to
$10^{13}$ G when $B_{\rm 0}\sim 15B_{\rm Q}$ (See
Eq.(20)). This energy also agrees with the observations of bursts.\\

The Fermi liquid of nucleons in a magnetar may be in metastable
state. A series of shifts in the deep crust closed to the Fermi
liquid of nucleons  can trigger the phase transition of the Fermi
liquid of nucleons from a metastable state to a stable state. The
free magnetic energy in the Fermi liquid  is released and it is far
greater than that in the crust because the magnetic induction and
the thickness of the Fermi liquid  is larger than the crust. The
actual size of the fermi liquid of nucleons in metastable state is
also problematical because of the unknowing configuration of
electrons and magnetic field distributions. However, all the Fermi
liquid of nucleons may evolve into metastable state simultaneously
when the deep crust is stable. Let the magnetic induction of the
Fermi liquid  be at the same order of magnitude as that in the deep
crust, and the thickness of the Fermi liquid be approximated by the
radius of the magnetar. We estimate that the energy released
is
\begin{equation}
E_{\rm flare}\sim 10^{44}(\frac{\delta B}{10^{13}})^{2}\ \ \  \ {\rm
ergs}.
\end{equation}
This energy agrees with those released in the GFs of some magnetars.\\
\maketitle
\section{Summary}
 We discussed the magnetization effects of the
relativistic degenerate electron gas in a neutron star. Having considered the
magnetic interaction among electrons, we found that the magnetic
systems may be unstable or metastable when the differential
susceptibility equals or exceeds 1. Under an ultra-strong magnetic
field, the magnetic domain structures in a magnetar can appear in the
solid crust, while the Fermi liquid of nucleons may be in metastable
state. The shearing stress acting on adjacent domains can result in
a series of shifts or fractures in the crust. The crust releases the
magnetic free energy, which corresponds to the bursts observed in
magnetars. Simultaneously, a series of shifts or fractures in the
deep crust closed to the Fermi liquid of nucleons  can
trigger the phase transition of the Fermi liquid  from
a metastable state to a stable state. The free magnetic energy in
the Fermi liquid of nucleons is released, which corresponds to the
GFs observed in some magnetars.

\section*{Acknowledgments}
We thank the anonymous referee for his/ her comments which
helped to improve the paper. ZJ thanks Prof. Anzhong Wang for correcting the
English language of the manuscript.
This work was supported in part by the National Science Foundation of China
(Grants Nos. 10963003, 11063002 and 11163005), Foundation of Huoyingdong under No. 121107,
Foundation of Ministry of Education under No. 211198 and the Doctor Foundation of
Xinjiang University (Grant No.BS110108).

\bibliographystyle{apj}
\bibliography{axp}

\begin{thebibliography}{32}
\expandafter\ifx\csname natexlab\endcsname\relax\def\natexlab#1{#1}\fi

\bibitem[{{Baym} \& {Pines}(1971)}]{Baym71}
{Baym}, G., \& {Pines}, D. 1971, Annals of Physics, 66, 816

\bibitem[{{Blandford} \& {Hernquist}(1982)}]{Blandford82}
{Blandford}, R.~D., \& {Hernquist}, L. 1982, Journal of Physics C Solid State
  Physics, 15, 6233

\bibitem[{{Canuto} \& {Chiu}(1968)}]{canuto68}
{Canuto}, V., \& {Chiu}, H.-Y. 1968, Physical Review, 173, 1229

\bibitem[{{Chudnovsky}(1981)}]{chudnovsky81}
{Chudnovsky}, E.~M. 1981, Journal of Physics A Mathematical General, 14, 2091

\bibitem[{{Cline} {et~al.}(1982){Cline}, {Desai}, {Teegarden}, {Evans},
  {Klebesadel}, {Laros}, {Barat}, {Hurley}, {Niel}, \& {Weisskopf}}]{Cline82}
{Cline}, T.~L., {et~al.} 1982, ApJL, 255, L45

\bibitem[{{Condon}(1966)}]{condon1966}
{Condon}, J.~H. 1966, Physical Review, 145, 526

\bibitem[{{Dingle}(1952)}]{dingle1952}
{Dingle}, R.~B. 1952, Royal Society of London Proceedings Series A, 211, 500

\bibitem[{{Duncan}(2004)}]{Duncan04}
{Duncan}, R.~C. 2004, in Cosmic explosions in three dimensions, ed.
  P.~{H{\"o}flich}, P.~{Kumar}, \& J.~C. {Wheeler}, 285

\bibitem[{{Duncan} \& {Thompson}(1992)}]{Duncan92}
{Duncan}, R.~C., \& {Thompson}, C. 1992, ApJ, 392, L9

\bibitem[{{Goldreich} \& {Reisenegger}(1992)}]{Goldreich92}
{Goldreich}, P., \& {Reisenegger}, A. 1992, ApJ, 395, 250

\bibitem[{{Hurley} {et~al.}(1999){Hurley}, {Cline}, {Mazets}, {Barthelmy},
  {Butterworth}, {Marshall}, {Palmer}, {Aptekar}, {Golenetskii}, {Il'Inskii},
  {Frederiks}, {McTiernan}, {Gold}, \& {Trombka}}]{Hurley99}
{Hurley}, K., {et~al.} 1999, Nature, 397, 41

\bibitem[{{Johnson} \& {Lippmann}(1949)}]{Jonhson49}
{Johnson}, M.~H., \& {Lippmann}, B.~A. 1949, Physical Review, 76, 828

\bibitem[{{Kondratyev}(2002)}]{Kondratyev02}
{Kondratyev}, V.~N. 2002, Physical Review Letters, 88, 221101

\bibitem[{{Lee} {et~al.}(1969){Lee}, {Canuto}, {Chiu}, \& {Chiuderi}}]{lee69}
{Lee}, H.~J., {Canuto}, V., {Chiu}, H.-Y., \& {Chiuderi}, C. 1969, Physical
  Review Letters, 23, 390

\bibitem[{{Lifshits} \& Kosevich(1956)}]{lifshitz1956}
{Lifshits}, E.~M., \& Kosevich, A.~M. 1956, Sov.Phys.-JETP, 2, 636

\bibitem[{{Mazets} {et~al.}(1999){Mazets}, {Cline}, {Aptekar'}, {Butterworth},
  {Frederiks}, {Golenetskii}, {Il'Inskii}, \& {Pal'Shin}}]{Mazets99}
{Mazets}, E.~P., {Cline}, T.~L., {Aptekar'}, R.~L., {Butterworth}, P.~S.,
  {Frederiks}, D.~D., {Golenetskii}, S.~V., {Il'Inskii}, V.~N., \& {Pal'Shin},
  V.~D. 1999, Astronomy Letters, 25, 635

\bibitem[{{Mazets} {et~al.}(1979){Mazets}, {Golentskii}, {Ilinskii}, {Aptekar},
  \& {Guryan}}]{Mazets79}
{Mazets}, E.~P., {Golentskii}, S.~V., {Ilinskii}, V.~N., {Aptekar}, R.~L., \&
  {Guryan}, I.~A. 1979, Nature, 282, 587

\bibitem[{{O'Connell} \& {Roussel}(1971)}]{oconnell71}
{O'Connell}, R.~F., \& {Roussel}, K.~M. 1971, Nature, 231, 32

\bibitem[{{Palmer} {et~al.}(2005){Palmer}, {Barthelmy}, {Gehrels}, {Kippen},
  {Cayton}, {Kouveliotou}, {Eichler}, {Wijers}, {Woods}, {Granot}, {Lyubarsky},
  {Ramirez-Ruiz}, {Barbier}, {Chester}, {Cummings}, {Fenimore}, {Finger},
  {Gaensler}, {Hullinger}, {Krimm}, {Markwardt}, {Nousek}, {Parsons}, {Patel},
  {Sakamoto}, {Sato}, {Suzuki}, \& {Tueller}}]{Palmer05}
{Palmer}, D.~M., {et~al.} 2005, Nature, 434, 1107

\bibitem[{{Pippard}(1963)}]{pippard1963}
{Pippard}, A.~B. 1963, Royal Society of London Proceedings Series A, 272, 192

\bibitem[{{Pippard}(1980)}]{Pippard80}
---. 1980, {in Electrons at the Fermi surface, ed M. Springford (Cambridge:
  Cambridge University press)}

\bibitem[{{Reichl}(1998)}]{Reichl98}
{Reichl}, L.~E. 1998, {A Modern Course in Statistical Physics, 2nd Edition}

\bibitem[{{Ruderman}(1991)}]{Ruderman91}
{Ruderman}, R. 1991, ApJ, 382, 576

\bibitem[{{Shoenberg}(1962)}]{Shoenberg1962}
{Shoenberg}, D. 1962, Royal Society of London Philosophical Transactions Series
  A, 255, 85

\bibitem[{{Shoenberg}(1984)}]{Shoenberg84}
---. 1984, {Magnetic oscillation in metals(Cambridge: Cambridge University
  press)}

\bibitem[{{Stella} {et~al.}(2005){Stella}, {Dall'Osso}, {Israel}, \&
  {Vecchio}}]{Stella05}
{Stella}, L., {Dall'Osso}, S., {Israel}, G.~L., \& {Vecchio}, A. 2005, ApJ,
  634, L165

\bibitem[{{Suh} \& {Mathews}(2010)}]{Suh10}
{Suh}, I.-S., \& {Mathews}, G.~J. 2010, ApJ, 717, 843

\bibitem[{{Terasawa} {et~al.}(2005){Terasawa}, {Tanaka}, {Takei}, {Kawai},
  {Yoshida}, {Nomoto}, {Yoshikawa}, {Saito}, {Kasaba}, {Takashima}, {Mukai},
  {Noda}, {Murakami}, {Watanabe}, {Muraki}, {Yokoyama}, \&
  {Hoshino}}]{Terasawa05}
{Terasawa}, T., {et~al.} 2005, Nature, 434, 1110

\bibitem[{{Thompson} \& {Duncan}(1995)}]{TD95}
{Thompson}, C., \& {Duncan}, R.~C. 1995, MNRAS, 275, 255

\bibitem[{{Thompson} \& {Duncan}(2001)}]{TD01}
---. 2001, ApJ, 561, 980

\bibitem[{{Visvanathan}(1962)}]{Visvanathan62}
{Visvanathan}, S. 1962, Physics of Fluids, 5, 701

\bibitem[{{Wilkes} \& {Ingraham}(1989)}]{Wilkes89}
{Wilkes}, J.~M., \& {Ingraham}, R.~L. 1989, ApJ, 344, 399

\end{thebibliography}

\end{document}